\documentclass[nohyper,notoc]{JHEP} 
\usepackage{epsfig,axodraw}

\def\beq{\begin{equation}}
\def\eeq{\end{equation}}
\def\bea{\begin{eqnarray}}
\def\eea{\end{eqnarray}}
\def\bq{\begin{quote}}
\def\eq{\end{quote}}
\def\gappeq{\mathrel{\rlap {\raise.5ex\hbox{$>$}}
{\lower.5ex\hbox{$\sim$}}}}
\def\lappeq{\mathrel{\rlap{\raise.5ex\hbox{$<$}}
{\lower.5ex\hbox{$\sim$}}}}

\def\msnu{m_{\tilde{\nu}}}

\def\Rpv{R_p \! \! \! \! \! \! /~}

\title{\large Neutrino masses in the $R_p$ violating MSSM }

\author{Sacha Davidson \\
 Theoretical Physics,  Oxford
University, 1 Keble Road, Oxford, OX1 3NP, United Kingdom}

\author{Marta Losada \\
 Centro de Investigaciones, 
Universidad Antonio Nari\~{n}o, Cll. 59 No. 37-71, Santa Fe de Bogot\'{a},
Colombia }

\abstract{
We compute one loop neutrino masses
in the R-parity violating Minimal Supersymmetric Model,
including the bilinear $\Rpv$ masses
 in the mass insertion approximation.
To the order
we calculate, our results are independent
of the Higgs-lepton basis choice. 
We have a variety of perturbative
parameters---gauge, yukawa and
trilinear couplings, and $R_p$ violating
masses. Their relative magnitudes
determine which diagrams are
relevant for neutrino mass
calculations. We
 find  new loop diagrams which
can be relevant and have 
frequently been neglected in the past. For
the Grossman-Haber 
neutral loop contribution to the neutrino mass matrix
we obtain explicit analytic results.}


\begin{document} 


\section{Introduction}

An interesting challenge in particle physics
is to construct a consistent model which can accomodate 
neutrino oscillations in order to explain atmospheric
and solar neutrino data \cite{SuperK}. Considering the results
of SuperKamiokande and all solar neutrinos experiments,
at least two different values of $\Delta m^2$ and
$\sin^2 2\theta$ are necessary to account for the
data. Neutrino masses cannot be generated in the Standard Model (SM),
so it  is necessary to consider extensions of the
SM \cite{neutrinos}. An interesting possibility for generating neutrino
masses are models in which lepton number violation occurs. 
In these models Majorana neutrino masses are generated by
interactions which violate lepton number in two units: $\Delta L=2$. 

Among the possible models we consider the Minimal Supersymmetric Standard Model
without imposing $R$-parity. The quantum number $R$ is defined as 
$R_p=(-1)^{L+3B+2S}$, where $L$, $B$, $S$ are the
  lepton and
 baryon number and the spin of the particle, respectively\cite{fayet}.
Imposing $R$-parity in a supersymmetric model has the advantage of
suppressing simultaneously the presence of baryon and lepton number violating
interactions, which are strictly constrained from proton decay experiments
and other low-energy processes \cite{dreiner}.
However, one can impose a less stringent constraint in which
all interactions conserve baryon number. Thus only lepton number violating
terms are present in the Lagrangian. These terms 
  allow left-handed
 neutrinos to obtain a
 Majorana mass, at tree level through mixing with the neutralinos,
 and through loop diagrams that violate lepton number 
(in two units)\cite{hall, Nilles, GH}.

In the SM, the Higgs and leptons have the same gauge quantum numbers.
However, they cannot mix  because the Higgs is 
a boson and the leptons are fermions. In a supersymmetric model
this distinction is removed, so the down-type
Higgs and sleptons can be assembled in a vector $L_J = (H_d, L_i)$ 
with $J:4..1 $.
With this notation, 
the superpotential for the  supersymmetric
SM with   $R_p$ violation  can be written as
\beq
W= \mu^J {H}_u  L_J + \lambda^{JK \ell} L_J L_K E^c_{\ell} + 
\lambda^{'Jpq} L_JQ_p D^c_q  + h_t^{pq} {H}_u Q_p U^c_q \label{S}
\eeq
The $R_p$ violating and
conserving coupling constants
have been assembled into vectors and matrices in $L_J$ space:
we call the usual $\mu$ parameter $\mu_4$,
and identify the usual $\epsilon_i =  \mu_i$, 
$\frac{1}{2}h_e^{jk} = \lambda^{4jk}$, and
$h_d^{pq} = \lambda^{' 4pq}$.
Lower case roman indices $i,j,k$ and $p,q$  are lepton and
quark generation indices. We frequently suppress the
capitalised indices, writing
$\vec{\mu} = ( \mu_4, \mu_3, \mu_2, \mu_1)$.

We also include possible $R_p$ violating couplings among
the soft SUSY breaking parameters, which can
be written as
\begin{eqnarray}
V_{soft} = & \frac{\tilde{m}_u^2}{2} H_u^{\dagger} H_u + \frac{1}{2}
 L^{J \dagger} [\tilde{m}^2_L]_{JK} L^K  + B^J H_u L_J   \nonumber \\ 
& + A^{ups} H_u Q_p U^c_s + 
     A^{Jps} L_J Q_p D^c_s +
    A^{JKl} L_J L_K E^c_l + h.c.~~~.  \label{soft}
\end{eqnarray}
Note that we have absorbed the superpotential parameters
 into the $A$ and $B$ terms; $e.g.$ we write $B^4 H_uH_d$
not $B^4 \mu^4 H_u H_d$ ~\footnote{We do this because $B_J$
is a vector---a one index object---in $\{L_J\}$ space.
From this perspective, giving it two indices can lead to confusion.}.
We abusively use capitals for superfields (as in (\ref{S})) and
for their scalar components. 

We have put the Higgs $H_d$ into a vector with the sleptons,
and combined the $R_p$ -violating with the $R_p$ conserving
couplings, because the lepton
number violation can be moved around the Lagrangian
by judiciously choosing which linear combination
of  hypercharge = -1  doublets  to identify as the
Higgs/higgsino, with the remaining doublets
being sleptons/leptons. 
However, this freedom
to redefine what violates $L$ 
is deceptive,  because phenomenologically we know that
the leptons are the mass eigenstate $e, \mu$ and $\tau$ ,
so we know what lepton number violation is.
Lepton number is defined in the
charged lepton mass eigenstate basis---the
freedom to choose which direction is the Higgs
in the Lagrangian just means that there is
not a unique interaction eigenstate basis.
There are two possible approaches to this fictitious
freedom; choose to work
in a Lagrangian basis that corresponds to the
mass eigenstate basis of the leptons,
or  construct combinations
of coupling constants that are independent
of the basis choice to parametrise the $R_p$
violation in the Lagrangian \cite{Nilles,sacha,Fer}. 
These invariant
measures of $R_p$ violation in the Lagrangian
are analogous  to Jarlskog invariants  
which parametrise  CP violation.

 The standard option  is to
work in a basis that corresponds approximately
to the mass eigenstate basis of the leptons. 
For instance, if one chooses the Higgs direction in
$L_J$ space to be parallel to $\mu_J$, then the 
additional bilinears in the superpotential $\mu_i$
will be zero. In this basis, the sneutrino vevs are
constrained to be small by the
neutrino masses, so this is approximately
the lepton mass eigenstate basis. Lepton number violation
among the fermion tree-level masses
in this basis is small
by construction, so it makes sense to neglect
the bilinear $R_p$ violation
 in setting constraints on  the trilinears,
as is commonly done (for a review, see {\it.e.g.} \cite{dreiner}.
For a careful analysis including the bilinears, see \cite{kongetal}.
Note that in many cases, the most stringent constraints on
the additional $R-$parity violating parameters
come from neutrino anomaly data \cite{everyone}.)

The aim of the ``basis-independent'' approach is
to construct
combinations of coupling constants that
are invariant under rotations in
$L_I$ space, in terms of which one can express
physical observables. By judiciously combining
coupling constants one can find ``invariants'' which
are zero if $R_p$ is conserved,  so these invariants parametrise
$R_p$ violation in a basis-independent way.
For instance,  consider the superpotential of equation (\ref{S})
in the one generation limit, $I:4..3$.
 It appears
to have two $R_p$ violating interactions:
$\mu_{3} H_u L$ and $\lambda' LQD^c$.  It is
well known that one of these can be rotated into
the other by mixing $H_d$ and $L$  \cite{hall}. If
\bea
H_d' = \frac{\mu_4}{\sqrt{\mu_4^2 + \mu_{3}^2}} H_d +  
        \frac{\mu_{3}}{\sqrt{\mu_4^2 + \mu_{3}^2}} L \nonumber \\
L' = \frac{\mu_{3}}{\sqrt{\mu_4^2 + \mu_{3}^2}} H_d -  \label{simple}
        \frac{\mu_4}{\sqrt{\mu_4^2 + \mu_{3}^2}} L ~~~,
\eea
then the Lagrangian expressed in terms of $H_d'$ and
$L'$ contains no $  H_u L'$ term. One could instead dispose 
 of the $\lambda' L QD^c$ term.
The coupling constant
combination that is invariant under basis redefinitions
in $(H_d,~L)$ space, zero if $R$ parity is conserved,
and non-zero if it is not is $\mu_4 \lambda' - h_d \mu_{3} =
( \mu_4, \mu_{3}) \wedge ( h_d, \lambda')$. For a more detailed
discussion of the approach followed in this
paper see {\it eg} \cite{sacha,Fer,DLR}.

 Basis-independent coupling constant combinations
which parametrise
the amount of $\Rpv$ can be constructed
in various ways. Their advantage over
the coupling constants  is
that they  $cannot$ be set to
zero by  a basis transformation.
So we define the following
invariants $\{ \delta \}$  which
reduce to the coupling constant
of the same indices in the basis
where the sneutrinos do not have vevs
(the $ \rightarrow$ is to what the
 $ \delta $ becomes in this basis).
\beq
\delta_{\mu}^i = \frac{ \vec{\mu} \cdot  \lambda^i \cdot  
\vec{ v}}{|  \vec{\mu}| m^e_i} 
\rightarrow 
\frac{\mu_i}{\mu_4}
\eeq
\beq
\delta_{\lambda'}^{ipq} = \frac{  \vec{\lambda}^{'pq}  \cdot 
\lambda^i  \cdot   \vec{v}}
{m^e_i} \rightarrow {\lambda}^{'ipq}
\eeq
\beq
\delta_{B}^i = \frac{  \vec{B} \cdot  \lambda^i  \cdot  \vec{v}}
{|  \vec{B}| m^e_i} \rightarrow \frac{B_i}{B_4}
\eeq
\beq
\delta_{\lambda}^{ijk} = \frac{ \vec{v}  \cdot \lambda^{i}
\lambda^{k} \lambda^j  \cdot  \vec{v}}
{m^e_i m^e_j} \rightarrow {\lambda}^{ijk}
\eeq
We have chosen the $Q$ and $D^c$ bases
to make $  \vec{v}   \cdot \lambda^{'pq}$ diagonal---that
is, to diagonalise the down-type quark mass matrix. We
assume that this is also the mass eigenstate basis
for the squarks. Similarly, we choose
the $E^c$ basis to make  $  \vec{v}  \cdot
\lambda^{i} \lambda^k  \cdot \vec{v} \propto 
\delta^{ik}$. Combined with our definition of the
lepton directions as $\hat{L}_i  =  \vec{v} \cdot \lambda^i/m_i^e$,
 ($m_i^e = | \vec{v} \cdot \lambda^i|$)
this means we diagonalise the charged lepton mass matrix
induced by the Higgs vev.
Here we are neglecting  
$R$-parity violating bilinears that mix  the charginos
and charged leptons; these masses will be included
as mass insertions in perturbation theory.

The  mass matrix of the neutralinos is composed by the 4
neutralino fields of the R-parity conservng MSSM and the 3
neutrino fields. In an interaction
eigenstate basis they are 
\beq
\chi_{int} = 
( \tilde{B}, \tilde{W}_3,
\tilde{h}_u,\tilde{h}_d, \nu_{\tau}, \nu_{\mu}, \nu_{e})
\eeq
where as previously discussed, in  $R_p \! \! \! \! \! \! /~$
models there is no unique interaction
eigenstate choice for the basis in $\tilde{h}_d$ and lepton
space. In this paper we define the basis where
the sneutrinos  do not have vacuum expectation values  to be
the interaction eigenstate basis.
We number the elements of $\chi$ in
{\it reverse} order, so 
$\chi_{int}^7 = \tilde{B}$, and   
 $(\chi_{int}^4,\chi_{int}^3,\chi_{int}^2, \chi_{int}^1) $ $=
(\tilde{h}_d, \nu_{\tau}, \nu_{\mu}, \nu_{e})$.
We use this numbering so that 
$(\chi^3,\chi^2, \chi^1) $
will be the mass eigenstate neutrinos.

In the $R_p$ conserving MSSM,
lepton number is conserved,  and the
first four fermions have majorana masses,
while the three neutrinos are massless.  
The mass matrix 
can be diagonalised
with a  matrix :
\beq
 \left[\begin{array}{cc}
   Z & 0 \\
   0 & I
\end{array} \right]
\eeq
where $Z$ is a $ 4 \times 4$ matrix
and $I$ is the $ 3\times 3$ identity matrix.

The purpose of this paper is to address and clarify the
issue of different  basis in $\Rpv$ neutrino mass
calculations. We
construct ``basis-independent'' estimates of
each loop diagram contributing to the
neutrino mass matrix. These estimates are
in an arbitrary Lagrangian basis. They lead us to include
the bilinear $\Rpv$ masses
in the mass insertion approximation. This
introduces a new diagram, and new contributions
from the usual diagrams. These new terms resolve
the basis related puzzles that arise
in  $\Rpv$ neutrino mass calculations
when the bilinears are neglected in the loops.
Clearly  neutrino masses 
should not depend on the basis they are computed in;
however,  if  the  $\Rpv$ masses  are neglected in
the loops, then what one is neglecting
depends on the basis. We will return
to this in the discussion at the end.
Having verified the irrelevance of performing 
basis independent calculations, 
we present estimates and the calculation
of the neutral loop in the basis where the sneutrinos do not have vacuum 
expectation values, because the results
are more compact than the basis independent estimates.

In section 2 we introduce our 
procedure for obtaining neutrino masses, and in section
3 we estimate the basis-independent contributions from all
loops. In section 4 
we  provide complete analytic expressions for
the neutral loop, and compare
with the exact result previously obtained in the literature\cite{DLR}. 
 The latter calculation is
done in the MSSM mass eigenstate basis. The
neutral graph is of interest as it has been analysed very
little in the literature and furthermore it can provide strong constraints
 on Higgs physics in this model.
In section 5 we discuss in more detail the issues related to the
basis choice and conclude.

\section{Perturbation theory with $\Rpv$ masses}  

Neutrino
masses are observationally known to be small,
so we can compute them  in perturbation
theory. There are two ways to include the $\Rpv$ bilinears in
one-loop calculations of neutrino masses.  One
approach is to diagonalise the tree level mass
matrices
for neutral and charged scalars and fermions, and 
 calculate the  one-loop contributions to the
neutrino mass matrix
using a tree-level 
 mass eigenstate basis for the propagators in the loop \cite{num}.
This is usually done numerically.
Note that for $\delta_{\mu} \neq 0$, $\delta_B \neq 0$, the
charged and neutral Higgses [MSSM neutralinos] mix with
the sleptons [leptons].
The alternative, which we follow here,
is to propagate particles in the MSSM mass
eigenstate basis, and include the $\Rpv$ masses  as 
mass insertions \cite{massins}
in perturbation theory. 
For
``sensible'' basis choices
that are close to the MSSM, 
the $\Rpv$ bilinear masses and couplings are small, so
including them as mass insertions should
be adequate.

The algorithm for computing neutrino masses, with
$\Rpv$ bilinears included in the mass
insertion approximation, consists
of the following
steps:
\begin{enumerate}
\item choose a basis where the $R_p$ violating
parameters are small. {\it eg} the basis where
the sneutrinos have no vevs, or where there are
no $\mu_i h_u \ell_i$ terms.
\item diagonalise the heavy neutralino $4 \times 4$
matrix.
\item calculate  the tree-level seesaw neutrino mass.

\item calculate loop contributions to 
the $3 \times 3$ neutrino mass matrix,
including the $\Rpv$ bilinear masses as
mass insertions.

\item diagonalise the resulting $3 \times 3$
neutrino mass matrix.
\end{enumerate}

It is well known that in the $\Rpv$ MSSM  only one of
the neutrinos acquires a mass at tree-level through
the see-saw mechanism due to the mixing of the
neutrinos and neutralinos.
We will focus on the (finite) loop  contributions to
neutrinos which are massless at tree level.
 If the tree-level
mass is non-zero, there are gauge loop
corrections which must be renormalised.
To avoid these, we for simplicity
neglect loop corrections to the
neutrino who is massive at tree level. 
We will address loops containing gauge bosons,
and issues of renormalisation,  in a
subsequent publication.
 If the
tree-level mass is smaller
than the loop contribution, it can be set to
zero ($\delta_{\mu} = 0$), and one-loop
masses can be calculated for
 three neutrinos.

We only consider one-loop
diagrams contributing to the neutrino
mass matrix---higher order gauge
loops may
be larger than
one loop diagrams with Yukawas/trilinears
at the vertices, but gauge
couplings have no flavour
indices so pure gauge loops will
give a mass aligned with that of
some lower order diagram. Two-loop
 diagrams with Higgs exchange rather
than mass insertions on the internal
lines should be suppressed by
an additional factor of $1/(16 \pi^2)$.
We neglect
the loop corrections to the neutralino sector and
to neutrino-neutralino mixing.
These would  contribute to the two loop neutrino mass matrix.
To compute the one-loop neutrino mass matrix,
we propagate tree level mass eigenstates
in the loop with tree level interactions.

In this paper we concentrate on step 4. We draw diagrams
and make basis-independent estimates in
an arbitrary Lagrangian
basis, which we assume satisfies step 1.
The advantage of this is that 
the basis-independent coupling constant combination
controlling the neutrino mass
can be read off these diagrams. However,
to get the correct dependence on MSSM
masses and mixing angles, we should
work in the MSSM mass eigenstate
basis. We do this for the neutral
loop in section 4, defining the Higgs to be
the direction in $L_I$ space that gets a vev.
 This basis choice in $L_J$ space is close to
the charged lepton mass eigenstate basis because
the misalignment between
$\vec{\mu}$ and $\vec{v}$ must be small,
so $\mu_i$ in this basis are small.
There are also no lepton number violating
D-term masses proportional to the sneutrino
vev. Since we are perturbing in lepton
number violating masses, this reduces the number
of interactions we must consider.
This is therefore a good basis from
which to do perturbation theory in
$\Rpv$ parameters.
We define our ``MSSM mass eigenstate basis''
(in an $\Rpv$ theory) by diagonalising
the mass matrices  with
all the $\Rpv$ parameters set to zero,
then reintroducing the  $\Rpv$ parameters
in the $L_J$ basis where the sneutrino
have no vevs.

We  have many parameters in which
we will be perturbing to compute loop
neutrino masses: Standard Model
Yukawa couplings $h$ and gauge couplings $g$,
of very different sizes, and 
bilinear and trilinear $\Rpv$
couplings of unknown size.
We parametrise the size of the $\Rpv$   
couplings in a basis-independent way
via the parameters $\delta$
introduced earlier. 
 We  would like to
order them in magnitude, so that
we  consistently
include  all contributions
greater than a certain size.

It is clear that $h \ll g$, and
that the Yukawas are known and of
different sizes.  We do not
know the size of the
various $\{\delta \} $; so
we consider three possibilities
\footnote{We have suppressed the family indices
for simplicity.}:
\begin{itemize}
\item {\bf A:} Assume   bilinear  $\Rpv$~ is small and
can be neglected  in the loops. This is the case if
 $ h_e\delta_{\lambda},  h_d\delta_{\lambda'} 
\gg g  \delta_{B},  g  \delta_{\mu}$.
One gets 
the usual loop diagrams  of figures  1a  and 1b.
\item  {\bf B:} Neglect $\delta_{\mu}$, 
but include the soft bilinear  $\Rpv$ terms ($B_i$
and $m^2_{4i}$) in perturbation theory.
This means  $ h_e \delta_{\lambda},  h_d \delta_{\lambda'}
 \sim g  \delta_{B}$, but 
 $ g \delta_{\mu} \ll  h_e \delta_{\lambda}, h_d \delta_{\lambda'}$.
This gives   the neutral Grossman-Haber loop
of figure 1c, and additional contributions 
to the usual loop of figure 1a.
\item   {\bf C:} 
Include all bilinear $\Rpv$ terms in perturbation theory.
There are additional contributions to diagrams
1a, 1b, and 1c, and a new diagram 1d.

\end{itemize}
 
Note that models with only bilinear $\Rpv$~ can
fall into any of these three categories. A model where
the misalignment
between the vev and the Yukawa couplings is
much greater than between the vev and
$\vec{\mu}$ would fall into case A.
Case C corresponds to a model
where the misalignment between
 the vev $\vec{v}$ and $\vec{\mu}$ is 
similar to the misalignment between
the Yukawa couplings and  $\vec{v}$.

\section{Estimates}

 The possible diagrams that contribute to the neutrino mass 
must contain two scalar-fermion-fermion couplings
at the vertices of the loop, and two $\Delta L = 1$ interactions.
The loop can contain  coloured and colour-singlet charged or neutral 
particles, and we can have either two gauge
couplings, one gauge and one Yukawa/trilinear coupling, or
two Yukawa/trilinear couplings at the vertices. 

The charged coloured loop is figure 1b. This
has Yukawa or trilinear couplings at the vertices,
where a higgsino or neutrino interacts with a quark
and a squark. Lepton number violation can be
due to the trilinears, or to
mass insertions
mixing the neutrino with a neutralino on the
external legs. 

In figure 1c we have the contribution from the neutral loop.  
At the vertices we must have two gauge couplings. 
It is not possible to have vertices with
combinations of gauge-Yukawa and Yukawa-Yukawa  couplings for a 
neutral loop, because the
vertex with the Yukawa coupling would involve an $E^c$.
This diagram has been previously considered in
\cite{GH}, and we will refer to it as
 the Grossman-Haber (GH) diagram.
Lepton number violation, $\Delta L = 2$, occurs
due to mass insertions  on
 the internal scalar propagator. 
 In the $<\tilde{\nu}> = 0$ basis,
we cannot have a neutral loop with 
one unit of lepton number violation  
on each  of the propagators
because if we turn the $\tilde{z}$ into a
$\nu$, we would also need  $ \partial_{\mu}
\tilde{\nu} \rightarrow Z_{\mu}$.
This is not possible,
because the Goldstone boson is in the direction
of the vevs, so has no  $ \tilde{\nu}$ component.

There are in addition two  charged loops, drawn in figures 1a and 1d.
Charged loops cannot have  two gauge
couplings at the incoming and outgoing vertices  because
$\Delta L = 2$ on a charged
line is forbidden by charge conservation.
It is not possible to have 
$\Delta L = 1$ on each line for
the same reason as for the neutral gauge
loop.  One gauge vertex will produce
a charged slepton
and a wino/chargino; if $\tilde{w}^+ \rightarrow \ell$
by a $\Delta L = 1$ mass, then
 $\partial^{\mu} \tilde{L}^+ \rightarrow W^{\mu}$
would be required on the bosonic line.
The  Goldstone boson has no $  \tilde{L}^+$
component in the  $<\tilde{\nu}> = 0$ basis.
For  charged
loops we can have two Yukawa/trilinear couplings at the vertices, which
is the canonical trilinear diagram of figures 1a.

It is important to note that charged loops
can have one gauge and one Yukawa coupling at the vertices. 
We show this contribution  in figure 1d.
To have one gauge and one Yukawa/trilinear coupling,
we need (charged) gaugino-lepton mixing on the fermion
line. This means we need $\delta_{\mu} \neq 0$,
so these diagrams only contribute 
when we make assumption C. We will discuss the
supersymmetric partner of diagram
1d, which could arise if $ W^{\mu}$ mixes with
 $\partial^{\mu} \tilde{E}^c$, in a subsequent
publication.

In the following subsections
we make basis-independent estimates of the
size of each loop, assuming that 
the Higgs vacuum expectation value, Higgs masses,
slepton masses and chargino/neutralino
masses are all of the same order $\sim m_{SUSY}$. We will
calculate 
the dependence of the neutral
loop  on the masses of the propagating
particles in section 4. We 
 also present  estimates  in the basis where the sneutrinos do not have a vev, $\langle\tilde{\nu}\rangle = 0$.
A more detailed discussion of the diagrams involving charged loops
will be given 
in a subsequent publication.

\subsection{$\delta_{\mu}, \delta_{B} \sim 0$: 
the canonical loop}

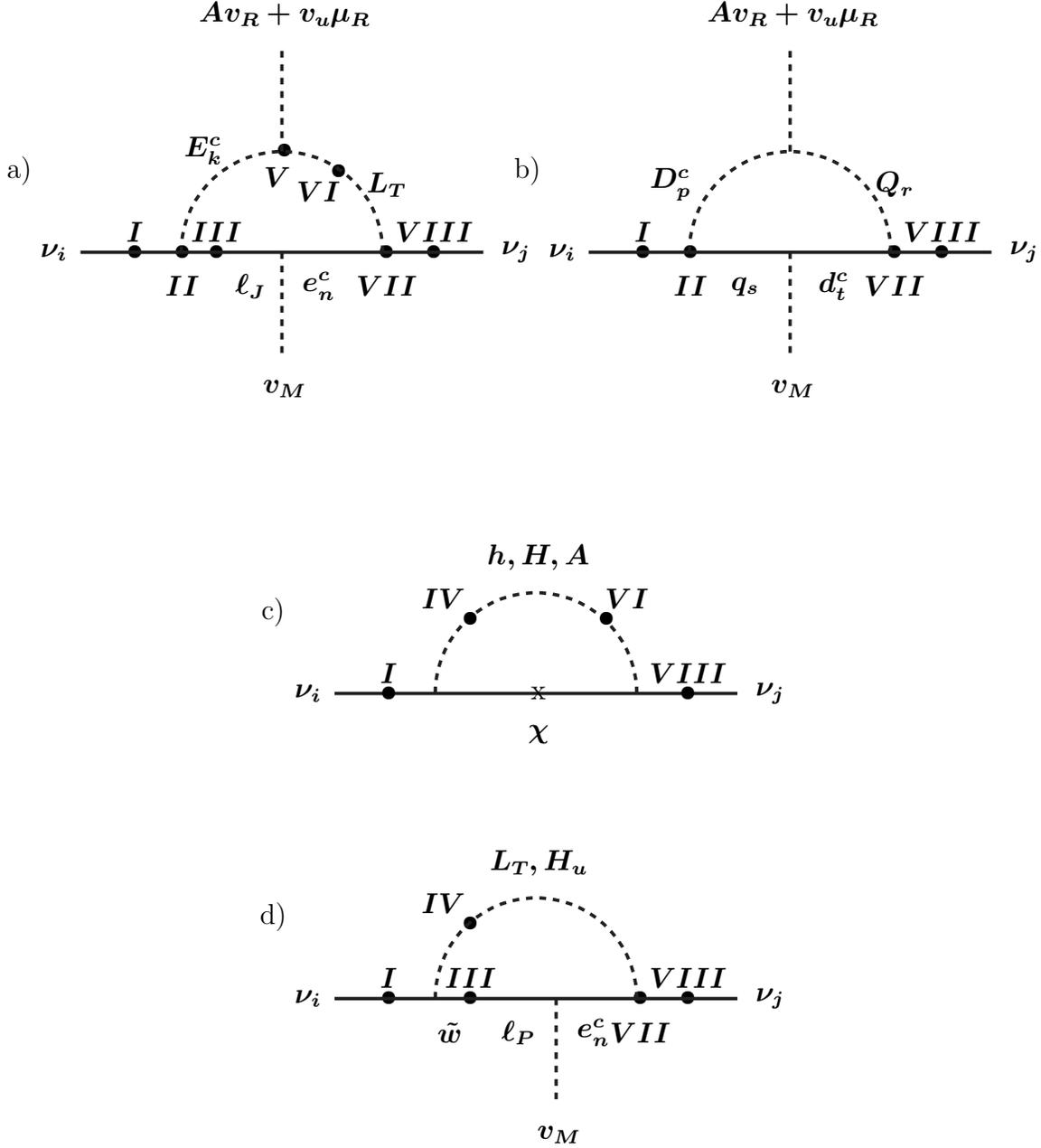
\begin{figure}[htb]
\unitlength1mm
\SetScale{2.8}
\begin{boldmath}
\begin{center}
\begin{picture}(60,60)(0,-20)
\Text(-9,12)[c]{a)}
\Text(8,0)[c]{$\bullet$}
\Text(8,3)[c]{$I$}
\Text(15,0)[c]{$\bullet$}
\Text(15,-5)[c]{$II$}
\Text(20,0)[c]{$\bullet$}
\Text(20,3)[c]{$III$}
\Text(45,0)[c]{$\bullet$}
\Text(45,-5)[c]{$VII$}
\Text(30,15)[c]{$\bullet$}
\Text(29,11)[c]{$V$}
\Text(52,0)[c]{$\bullet$}
\Text(52,3)[c]{$VIII$}
\Text(38,12)[c]{$\bullet$}
\Text(35,9)[c]{$VI$}
\Line(0,0)(15,0)
\Line(45,0)(15,0)
\Line(60,0)(45,0)
\DashCArc(30,0)(15,0,180){1}
\Text(-2,0)[r]{$\nu_i$}
\Text(62,0)[l]{$\nu_j$}
\Text(18,15)[c]{$E^c_k$}
\Text(45,10)[c]{$L_T$}
\Text(25,-5)[c]{$\ell_J$}
\Text(35,-5)[c]{$e^c_n$}
\Text(30,-20)[c]{$v_M$}
\DashLine(30,0)(30,-15){1}
\Text(30,35)[c]{$A v_R +  v_u \mu_R$}
\DashLine(30,15)(30,30){1}
\end{picture}
\hspace{12mm}
%
\begin{picture}(60,60)(0,-20)
\Text(-9,12)[c]{b)}
\Text(8,0)[c]{$\bullet$}
\Text(8,3)[c]{$I$}
\Text(15,0)[c]{$\bullet$}
\Text(15,-5)[c]{$II$}
\Text(45,0)[c]{$\bullet$}
\Text(45,-5)[c]{$VII$}
\Text(52,0)[c]{$\bullet$}
\Text(52,3)[c]{$VIII$}
\Line(0,0)(15,0)
\Line(45,0)(15,0)
\Line(60,0)(45,0)
\DashCArc(30,0)(15,0,180){1}
\Text(-2,0)[r]{$\nu_i$}
\Text(62,0)[l]{$\nu_j$}
\Text(12,10)[c]{$D^c_p$}
\Text(45,10)[c]{$Q_r$}
\Text(23,-5)[c]{$q_s$}
\Text(36,-5)[c]{$d^c_t$}
\Text(30,-20)[c]{$v_M$}
\DashLine(30,0)(30,-15){1}
\Text(30,35)[c]{$A v_R +  v_u \mu_R$}
\DashLine(30,15)(30,30){1}
\end{picture}
\end{center}
%
\begin{center}
\begin{picture}(60,40)(0,0)
\Text(-9,12)[c]{c)}
\Line(0,0)(15,0)
\Line(45,0)(15,0)
\Line(60,0)(45,0)
\DashCArc(30,0)(15,0,180){1}
\Text(8,0)[c]{$\bullet$}
\Text(8,3)[c]{$I$}
\Text(20,11)[c]{$\bullet$}
\Text(16,14)[c]{$IV$}
\Text(52,0)[c]{$\bullet$}
\Text(52,3)[c]{$VIII$}
\Text(40,11)[c]{$\bullet$}
\Text(43,14)[c]{$VI$}
\Text(-2,0)[r]{$\nu_i$}
\Text(62,0)[l]{$\nu_j$}
\Text(30,20)[c]{$h,H,A$}
\Text(30,0)[c]{x}
\Text(30,-6)[c]{$\chi$}
\end{picture}
\end{center}
\end{boldmath}
\begin{boldmath}
\begin{center}
\begin{picture}(60,60)(0,-20)
\Text(8,0)[c]{$\bullet$}
\Text(8,3)[c]{$I$}
\Text(20,0)[c]{$\bullet$}
\Text(20,3)[c]{$III$}
\Text(45,0)[c]{$\bullet$}
\Text(45,-5)[c]{$VII$}
\Text(52,0)[c]{$\bullet$}
\Text(52,3)[c]{$VIII$}
\Text(20,11)[c]{$\bullet$}
\Text(16,14)[c]{$IV$}
\Line(0,0)(15,0)
\Line(45,0)(15,0)
\Line(60,0)(45,0)
\DashCArc(30,0)(15,0,180){1}
\Text(-9,12)[c]{d)}
\Text(-2,0)[r]{$\nu_i$}
\Text(62,0)[l]{$\nu_j$}
\Text(30,20)[c]{$L_T, H_u$}
\Text(17,-5)[c]{$\tilde{w}$}
\Text(27,-5)[c]{$\ell_P$}
\Text(38,-5)[c]{$e^c_n$}
\Text(33,-20)[c]{$v_M$}
\DashLine(33,0)(33,-15){1}
\end{picture}
\end{center}
\end{boldmath}
\caption{Schematic 
representation of one-loop diagrams
contributing to neutrino masses,
in a Lagrangian basis. The blobs
indicate possible positions for $\Rpv$ interactions, which
can be 
trilinears (at positions II and VII) or mass insertions.
The misalignment between $\vec{\mu}$ and
$\vec{v}$ allows a mass insertion on
the lepton/higgsino lines (at points
I, III, or VIII) and at the $A$-term
on the scalar line (position V). The
soft $\Rpv$ masses appear as mass insertions
at positions VI and IV on
the scalar line. 
Figure a) is the charged loop with trilinear couplings
$\lambda$ (or $h_e$)  at the vertices.
Figure b)  is the  coloured loop with trilinear $\lambda'$ or
yukawa $h_b$  couplings.
Figure c) is the neutral  loop 
with two gauge couplings
(this diagram is drawn in the MSSM
mass eigenstate basis.), and 
figure d) is the charged loop
 with one gauge and and  a Yukawa coupling. This diagram is relevant
if gauginos mix with  charged leptons---that is if
$\delta_{\mu} \neq 0$.} 
\label{f1}
\end{figure}

Figures 1a and 1b  contain the usual diagrams considered in the literature
contributing to one loop neutrino masses when we allow the lepton
number violation to occur only at point II and VII. We neglect
the $\Rpv$ bilinear masses in this
section, so the only available sources
of $\Rpv$ are the trilinear couplings at
the vertices. Lines 1,2 of table 1  list 
 the basis-independent
 coupling constant combinations to which these diagrams
are proportional and  estimate the contribution to the
neutrino mass  matrix.
Note that on both the scalar and the fermion lines
the particle must flip between
doublet and singlet via an interaction with
the Higgs/slepton vev. We represent this as
a mass insertion of the vev to
assist in writing basis-independent estimates.

To get the estimate in column 4 of
table 1, we assume that
we have some incident $\nu_i$, in some
choice of basis.  At vertex $II$, we have
a trilinear coupling $\lambda^{iJk}$, where
in a generic basis we allow $J:1..4$ and
$k:1..3$. 
A slepton $E_k^c$ propagates
to the top of the loop,
where a trilinear mass insertion
$A^{RTk} v_R + \lambda^{RTk} \mu_R v_u$
turns the $E_k^c$ into
an $L_T$. We again generically allow
$T:4..1$---this means we have not 
chosen the singlet charged lepton basis.
 $L_T$ propagates to the vertex $VII$. Now
following the fermion line down, we propagate
an $\ell_J$ to the bottom of the loop,
flip it into an SU(2) singlet via the mass
$v_M \lambda^{JMn}$, and
propagate $e_n^c$ to vertex
$VII$, where we put a 
trilinear $ \lambda^{Tjn}$.

Thus, the contribution to the  neutrino mass matrix is
\beq
m_{ij} \propto \frac{ \lambda^{iJk}\lambda^{Tjn}}{16 \pi^2 m_{SUSY}^2} 
 \lambda^{JMn} v_M  [ (\lambda A)^{RTk} v_R +  v_u \mu_R \lambda^{RTk} ],
\label{lam}
\eeq
where all repeated indices are summed.
In basis where sneutrinos do not have a  vev, and assuming 
that $ (\lambda A)^{RTk}= A \lambda^{RTk}$
\footnote{This avoids flavour violation
among the sleptons. Additional diagrams would appear when one
allows flavour violation in the slepton sector \cite{KONG}.}
this is approximately 
\beq
m_{ij} \sim \frac{ \delta_{\lambda}^{ink} 
\delta_{\lambda}^{jkn}}{16 \pi^2} 
\frac{ m^e_{n} m^e_{k} }{ m_{SUSY}},
\label{lam:nu=0}
\eeq
where $ m^e_{n} = |\vec{v} \cdot \lambda^n|$ 
are the charged lepton masses generated by the Higgs vev,
for $n: 1..3$.
We can compare this expression to the more correct formula given by 
\beq
m_{ij} \sim \left \{ \lambda^{ink} \lambda^{jkn} m^e_{n} 
  m^e_{k} ( A + \mu \tan \beta )\right\}
\frac{f(L_{k})}{16 \pi^2 m^2_{E_k^c}},
\eeq 
where $L_{k}  =   m^2_{L^k}/m^2_{E_k^c}$ 
\beq
f(x) = \frac{\ln x }{  1 -x},
\eeq
and $ \mu = |\vec{\mu}|$.

Similarly, 
figure 1b with $\lambda'$ couplings produces a neutrino mass matrix
\beq
m_{ij} \propto  \frac{3 \lambda^{'isp}
 \lambda^{'jrt}}{16 \pi^2m_{SUSY}^2}  
\lambda^{'Mst} v_M ( (\lambda A)^{'Rrp} v_R +
 v_u \mu_R  \lambda^{'Rrp}).
\eeq
Recall that we chose the quark basis to diagonalise
the mass matrix $[ \vec{v} \cdot \lambda']^{pq}$. 
The exact result is
\beq
m_{ij} \sim 3 \lambda^{'ips}\lambda^{'jsp} m^d_{p} 
  m^d_{s}(A + \mu \tan \beta)\frac{f(Q_{p})}{16 \pi^2 m^2_{D_p^c}},
\eeq
where  $Q_{p}  =   m^2_{Q^p}/m^2_{D_p^c}$.

\subsection{$\delta_{\mu} \sim 0$,
 $  \delta_{B}\neq 0$: the GH loop}

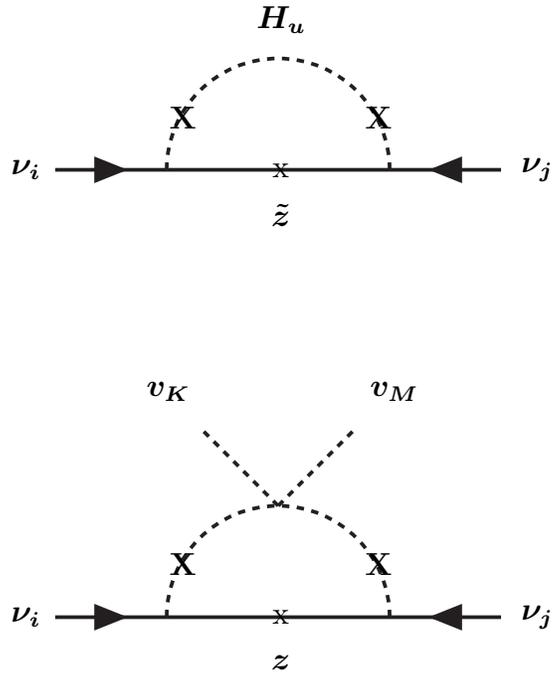
\begin{figure}[ht]
\unitlength1mm
\SetScale{2.8}
\begin{boldmath}
\begin{center}
\begin{picture}(60,40)(0,0)
\ArrowLine(0,0)(15,0)
\Line(45,0)(15,0)
\ArrowLine(60,0)(45,0)
\DashCArc(30,0)(15,0,180){1}
\Text(-2,0)[r]{$\nu_i$}
\Text(62,0)[l]{$\nu_j$}
\Text(17,7)[c]{{\bf X}}
\Text(43,7)[c]{{\bf X}}
\Text(30,20)[c]{$H_u$}
\Text(30,0)[c]{x}
\Text(30,-6)[c]{$\tilde{z}$}
\end{picture}
\end{center}
\vspace{10mm}
\begin{center}
\begin{picture}(60,40)(0,0)
\ArrowLine(0,0)(15,0)
\Line(45,0)(15,0)
\ArrowLine(60,0)(45,0)
\DashCArc(30,0)(15,0,180){1}
\Text(-2,0)[r]{$\nu_i$}
\Text(62,0)[l]{$\nu_j$}
\Text(17,7)[c]{{\bf X}}
\Text(43,7)[c]{{\bf X}}
\DashLine(30,15)(20,25){1}
\DashLine(30,15)(40,25){1}
\Text(15,30)[c]{$ v_K$}
\Text(45,30)[c]{$ v_M $}
\Text(30,0)[c]{x}
\Text(30,-6)[c]{$z$}
\end{picture}
\end{center}
\end{boldmath}
\vspace{10mm}
\caption{The Grossman-Haber  contribution
in interaction eigenstate basis.
The scalars leaving the neutrino
vertices are sneutrinos. The crosses
on the scalar line are $B_i$,  $B_j$, ${m}^2_{4i} $ or  ${m}^2_{4j}$. 
\label{f2a} }
\end{figure}

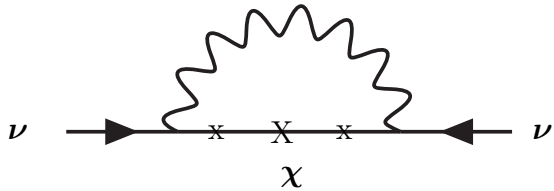
\begin{figure}[ht]
\unitlength1mm
\SetScale{2.8}
\begin{boldmath}
\begin{center}
\begin{picture}(60,40)(0,0)
\ArrowLine(0,0)(15,0)
\Line(45,0)(15,0)
\ArrowLine(60,0)(45,0)
\PhotonArc(30,0)(15,0,180){-2}{8}
\Text(-5,0)[r]{$\nu$}
\Text(62,0)[l]{$\nu$}
\Text(20,0)[c]{x}
\Text(28,0)[c]{{ X}}
\Text(37,0)[c]{x}
\Text(30,-6)[c]{$\chi$}
\end{picture}
\end{center}
\end{boldmath}
\vspace{10mm}
\caption{Feynman diagram showing the gauge loop correction to
the tree-level neutrino mass in MSSM
neutralino mass eigenstate basis. This is the supersymmetric
partner of the GH loop. The small ``$x$''s on
the internal fermion line correspond to
the interaction $\mu_i \nu_i h_u$. The
large $X$ is a neutralino mass. There
would be additional diagrams in a basis
where $<\tilde{\nu}> \neq 0$.}
\label{f2bb}
\end{figure}

As mentioned above, there can be   a loop contribution  of order $g^2$
to the mass of the neutrinos
that are massless at tree level
from the diagram of figure 1c.
This has been discussed in detail be
Grossman and Haber \cite{GH} (see
also \cite{Hetal,DLR}). The soft
SUSY breaking terms $B_k$ and 
${m}^2_{4k}$, and the sneutrino vev, 
mix the Higgses with the sleptons.
 If the soft
masses are universal and  $B_k = B \mu_k $,
diagram 1c  will be a loop correction
to the mass of the neutrino that is
massive at tree level. In
figure \ref{f2bb} we show the supersymmetric partner of this diagram which gives a
small contribution to the mass of the neutrino which is massive at tree level.  However if
$\vec{B}$ is misaligned with respect to
both $\vec{v}$ and $\vec{\mu}$, 
the diagram of figure 1c will give a mass also to 
the neutrinos that are massless 
at tree level.

The diagram in fig.1c is in the 
``MSSM mass eigenstate basis'',where
the sneutrino does not have a vev. The sneutrinos
mix with the neutral CP-even 
Higgses $h, H$ and the CP-odd Higgs $A$  through 
 $B_k$ and ${m}^2_{4k}$.
In Section 4, we calculate
this diagram in the mass insertion
approximation, and show
that we get the same answer as by
expanding the exact result of \cite{DLR}.
We  also show the contributions to this diagram
in an arbitrary  basis 
in figure  \ref{f2a}, 
because this  facilitates
basis-independent estimates.
There is an additional diagram (not drawn)
combining those of figure \ref{f2a}, with
 one unit of lepton number violation from
$B$ and one from $\tilde{m}^2$.
In an arbitrary interaction eigenstate basis,
the sneutrino could have a vev,
which induces lepton number violating masses through
the D-terms. This would
correspond to figure  \ref{f2a}
with diagonal $\tilde{m}^2$ and $L$ violation
from the sneutrino vevs.

Figure \ref{f2a} and
the diagrams with one unit of lepton number violation from
$B$ and one from $\tilde{m}^2$
 contribute to the  neutrino mass matrix by
\beq
m_{ij} \propto  g^{2} (v_uB_{i} + {m}^2_{iK} v^K)
(v_u B_{j} +  v^M {m}^2_{Mj}).
\eeq

We can write $v_J[m_L]^2_{Ji}$ in terms of  $B_{i}$
 using 
the minimisation conditions for the Higgs
potential \cite{DLR}:
\beq
v_u B_J + (\delta_{JK} m_Z^2 \cos 2 \beta + [m_L]^2_{JK}) v^K = 0,
\label{12}
\eeq
where at tree level $ [m_L]^2_{KJ} = \tilde{m}^2_{JK} + \mu_{J} \mu_{K}$,
and $B$ and $ \tilde{m}^2$ are defined in equation (\ref{soft}).

In Section 4,  we calculate  diagram 
1c,  and find that 
(in the lepton flavour basis 
where the sneutrino has no vev) it is approximately
\beq
m_{ij} \sim \frac{g^{2} \delta_B^{i}
 \delta_B^{j}}{64 \pi^{2} } m_{\chi}.
\eeq

Note that we neglect  the supersymmetric partner
of these diagrams, figure \ref{f2bb}, because it
is a small  correction to the tree-level
mass.  As explained above, the GH  diagram gives a mass to
a different combination of neutrinos
if $B_J$ is not exactly aligned with
$\mu_J$, that is $B_J \neq B \mu_J$.
So in the basis where
the sneutrino has no vev,
 we are only interested
in the component of $B_i$ which is
orthogonal to $\mu_i$. 
Writing 
\beq
B^i - \frac{B^i \mu_i}{\sum_i \mu_i^2} \mu^i \equiv B_{\perp}^i
= B^i - B^i_{\parallel},
\label{14}
\eeq
we find
\beq
m_{ij} \sim \frac{g^{2} (\delta_{B_{\perp}}^{i}
 \delta_{B_{\parallel}}^{j}  + \delta_{B_{\parallel}}^{i}
\delta_{B_{\perp}}^{j} 
+\delta_{B_{\perp}}^{i}\delta_{B_{\perp}}^{j} )}
{64 \pi^{2} } m_{\chi}.
\eeq
The component of  $B_i$
parallel to  $\mu_i$ contributes
to the mass of the neutrino
that is massive at tree level.
We also neglect 
the diagram $\sim \delta_{\mu}^i\delta_{\mu}^j h_t^4$.
This comes from mixing the neutrino with  the up-type higgsino, $\tilde{h}_u$,
which has a Majorana mass from a loop like
figure 1b with tops and stops instead of
bottoms/sbottoms.

There is an additional contribution
to the neutrino mass matrix
in the case that $ h \delta_B \sim \delta_{\lambda}$.
This comes from the diagram
of figure 1a,  with an 
$L$ violating mass insertion 
at point VI.  One unit
of  lepton number violation comes
from a trilinear, and another from
$B_k$ or $m^2_{4k}$. We use
the minimisation
condition for the Higgs/slepton potential:
$ - \tan \beta B_k = \tilde{m}^2_{4k}$
(in the $< \tilde{\nu}>  = 0 $ basis) \cite{DLR}, 
to get the neutrino mass estimate in table 1.

\subsection{ Contributions when $\delta_{\mu} \neq 0$,
 $ \delta_{B}\neq 0$}

In this section, we include $L$ violating mass insertions
on all possible propagators for diagrams with charged particles in
the loops. Consider
first the diagram of figure 1a. On the
scalar propagator there
can be a mass insertion $\delta_B$ discussed
at the end of the previous section,  or  $\delta_{\mu}$
(the latter appears in the off-diagonal mass term mixing the
$E^c$ with $L$ or $H$). Both external neutrino legs
and the internal lepton line can have mass insertions
$\delta_{\mu}$. Each
of these possibilities is
represented by a blob on the diagrams
of figure 1. The diagram is  shorthand for a series
of loops. 
The possibilities
are listed in table 1, along with the
corresponding estimates for $m_{\nu}$.

The diagrams must have two units
of same-sign lepton number violation
to generate a neutrino mass. 
Diagrams with $\Delta L = + 1$ and
$\Delta L = -1$ would contribute
to wavefunction renormalisation.
In  diagram 1a, this implies that we
need one unit of lepton number violation from
I--III and
one from IV --VIII.
 Schematically
we expect terms of order
\beq
(\delta_{\mu}+\delta_{\lambda}+
\delta_{\mu})(
\delta_{\mu}+\delta_{B}+\delta_{\lambda}+
\delta_{\mu})
\eeq
where the first parenthese corresponds to the blobs
I--III and the second to the blobs IV--VIII.
 We  neglect for convenience in
the table the $m^2_{4k}$ insertions because these
can be rewritten in terms of $B_i$ using
the minimisation conditions \cite{DLR}. The contributions
 of order $\delta_{\lambda}^2$, $\delta_{\lambda'}^2$, $\delta_{\lambda}\delta_{B}$
have already been discussed, the corresponding estimates
are given on the first, second and fifth lines of table 1, respectively.
  We briefly discuss here
how we calculated the results presented in table 1 for the
new diagrams which we consider.

We first include the $\delta_{\mu}$
perturbations on the external lines at points I and VIII. 
These will be present for both $\lambda$
(figure 1a) and $\lambda'$ (figure 1b) and correspond to expressions on
lines 4 and 6 of table 1. 
 We
neglect the diagram with mass insertions on both legs
because this will be proportional to the
tree level mass $\sim \mu_i \mu_j/m_{\chi}$.
We can estimate the diagram with lepton number violation
at points I and VII as follows.
In the $<\tilde{\nu}_k> = v_k = 0$ basis ($k: 1..3$),
the incident neutrino $\nu_i$ will be in
the direction $(\hat{\nu}_i)^J = v_M \lambda^{MJi}/m^e_i$. 
The mass insertion on the external leg is $\mu_i 
= \vec{v} \cdot \lambda^i \cdot \vec{\mu}/m^e_i \propto 
\delta_{\mu}^i$, and allows the incident neutrino
to turn into a Higgsino $h_u$. In an arbitrary basis,
where $ v_i \neq 0$ is allowed,
the mass insertion will be the $\mu_i v_4 - v_i \mu_4
\propto \vec{\mu} \cdot \lambda^i \cdot \vec{v}$. 
Continuing along the incident line, in the basis
where the sneutrinos do not have vevs, the $h_u$
can turn into an $h_d$ via $\mu_4$, and at the
vertex $h_d$ interacts with $E^c$ and $\ell$ via
a Yukawa coupling.  In an arbitrary basis, the neutralino after
the mass insertion of
$\mu_i$ or $v_i$ will be an $h_u$ or
a gaugino.  These can respectively
be turned into the linear combination of $\ell_I$
corresponding to the $h_d$  by $\mu_I$ or
$v_I$. These vectors are misaligned
by an amount $\delta_{\mu}$, which we can neglect
here because we have  one unit of $L$ violation
from the mass insertion. We assume
$|\vec{v}| \sim |\vec{\mu}| \sim m_{SUSY}$.
In column  four of table 1,
we therefore write the vertex as
$\mu_I \lambda^{'Ipq} [\mu_I \lambda^{IJk}]$
for the squark [charged] loop. The remainder
of the diagram is the same as lines 1 and 2 of
the table.

Additional terms of order $\delta_{\mu}(\delta_{\mu} + \delta_{\lambda})$
are obtained when we include the misalignment  between
$\vec{\mu}$ and $\vec{v}$ 
at point V of diagram 1a, giving the contributions on lines 7-9 of table 1.. We can express
\beq
\vec{\mu} = \frac{\vec{\mu} \cdot \vec{v}}{v^2} \vec{v} +
\frac{\vec{\mu} \cdot \lambda^{\ell} \cdot \vec{v}}{m_{\ell}^2}
 \vec{v} \cdot \lambda^{\ell},
\eeq
or alternatively the  part of 
$(v_R A + \mu_R v_u)\lambda^{RTk}$ that is misaligned
with respect to $  v_R \lambda^{RTk}$ can be written as
\beq
 v_u \frac{ \mu_R \lambda^{RP \ell} v_P  v_Q  \lambda^{QS \ell} \lambda^{ST k}}
{ m_{ \ell}^2}  ~~~~.
\eeq

Finally there are corrections proportional to $\delta_{\mu}$
from $\mu_i$ mass insertions on the internal lines of
figure 1a. 
We assume $ 
  \vec{v}_J \lambda^{JLi} \lambda^{LMk}   \vec{v}_M =\delta^{ik} m_i m_k$
(because $v - \mu$ misalignment is higher order effect)
to get the results listed in lines 9-13 of table 1.

If $\delta_{\mu} \neq 0$, the $\tilde{w}$ mixes
with the charged leptons, so there
are   loop diagrams
with a gauge coupling at one end and
a Yukawa coupling at the other, $e.g.$ 
see figure 1d.
These gauge $\times$ yukawa diagrams
can only appear if the gaugino mixes
with the lepton on the internal line,
so they are proportional to $\delta_{\mu}$.
The other unit of $L$ violation
must be due to $\delta_{\mu}$ (at points I or
VIII)  or $\delta_{B}$ (at point IV). These contributions
are estimated in lines 14-16 of table 1.
There is no contribution with
a trilinear $\lambda$ at the vertex
VII because it gives $[m_{\nu}^{ij}] 
\sim \lambda^{ijk}$, and $\lambda^{ijk}$
is anti-symmetric on indices $ij$
whereas the mass matrix is symmetric.

\FIGURE[ht]{
\epsfig{file=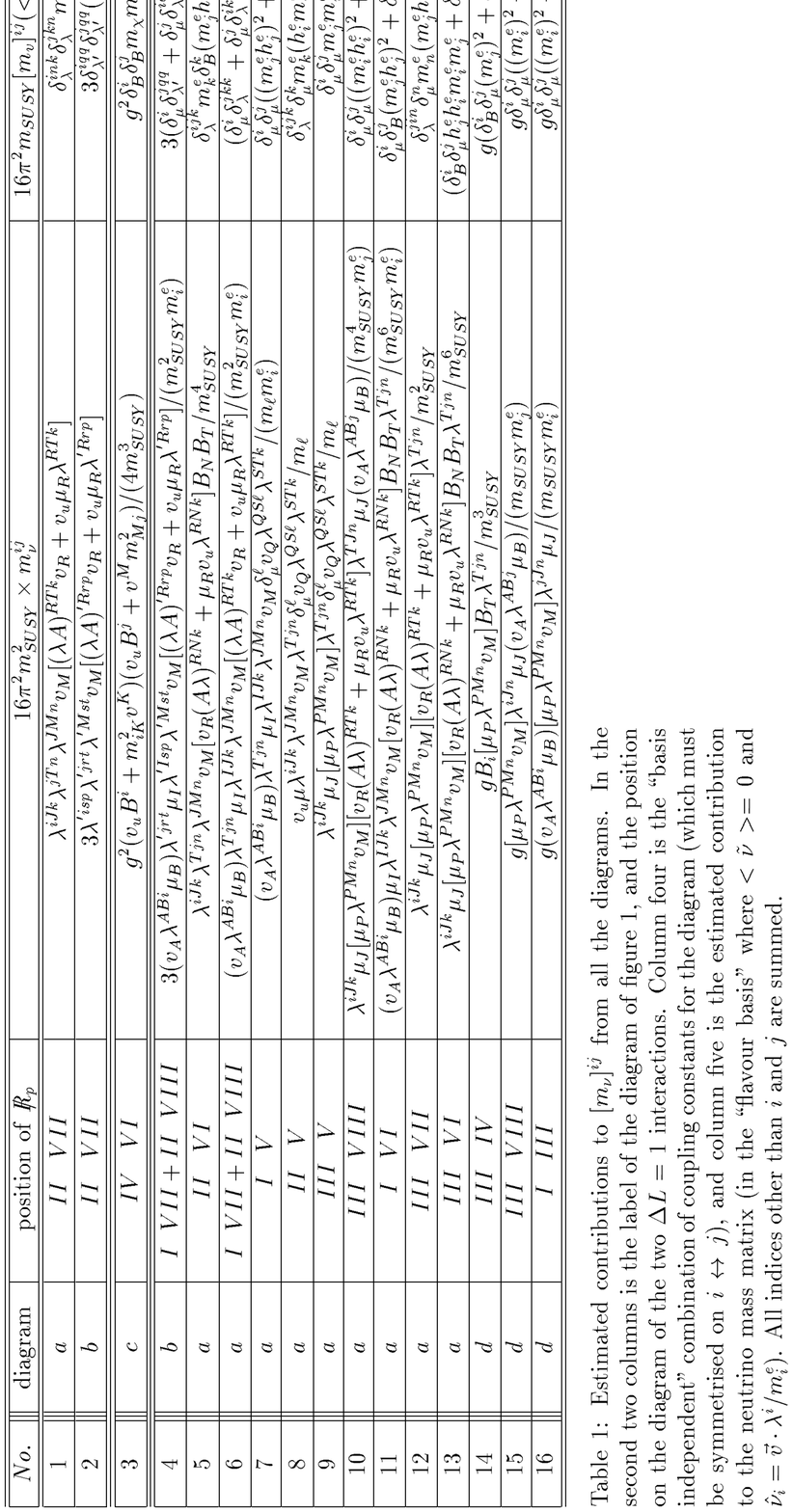,height=25cm}}

\section{ The  GH loop in the MSSM mass eigenstate basis }

In this section, we calculate  the one loop neutrino mass 
induced by slepton-higgs mixing (figure 1c). We
first do this using mass insertions the MSSM mass
eigenstate basis, then compare to our
previous exact (and basis independent)
result in one generation \cite{DLR}. The
two calculations agree,  but differ
slightly from the original calculation
of \cite{GH}. 

In the 
$<\tilde{\nu}> =0$ basis
the two units of lepton number violation come from
 $B_{i}$ and
$[m_L]^2_{4i}$
(the crosses on the scalar propagators
in figure 1c),
 which mix the sneutrino
with the Higgses.  The scalars that leave
the vertex where the neutrinos  go in are
sneutrinos.
We can write  $[m_L]^2_{4i} = -\tan \beta B_i$,
using equation (\ref{12}) in the basis where
$<\tilde{\nu}> = 0$.

 The amplitude
for this diagram connecting 
a neutrino $\nu_i$ to  $\nu_j$ is
\begin{eqnarray}
{\cal A}_{ij}& = & B_i B_j  \frac{g^2}{4} \sum_{\chi_{\alpha}}  
m_{\chi_{\alpha}} (Z_{{\alpha}2} - Z_{{\alpha}1} g'/g)^2 \times \nonumber\\
&&   \left\{ (  \cos^2 \alpha 
 + 2 \tan \beta \cos \alpha \sin \alpha 
 + \tan^2 \beta \sin^2  \alpha)
I(m_h^2, m_{\tilde{\nu}}^2, m_{\chi_{\alpha}}^2)  \right. \nonumber\\ && + 
  ( \sin^2 \alpha 
 - 2 \tan \beta \cos \alpha \sin \alpha 
 + \tan^2 \beta \cos^2  \alpha)
I(m_H^2, m_{\tilde{\nu}}^2, m_{\chi_{\alpha}}^2) \nonumber\\ &&
-  \left.  (  \cos^2 \beta 
 + 2 \tan \beta \cos \beta \sin \beta 
 + \tan^2 \beta \sin^2  \beta)
I(m_A^2, m_{\tilde{\nu}}^2, m_{\chi_{\alpha}}^2) \right\}.
\end{eqnarray}
Recall $\{ \chi_{\alpha} \}$ for $\alpha: 7..4$ are
the MSSM neutralinos, and  $h,H,$ and $A$ are the usual
MSSM fields:
\begin{eqnarray}
h &=& \cos \alpha H_u^R - \sin \alpha H_d^R \\
H &=& \sin \alpha H_u^R + \cos \alpha H_d^R \\
A & = & \cos \beta H_u^I - \sin \beta H_d^I,
\end{eqnarray}
($ H_u^R$ and $ H_u^I$ are the real and imaginary parts of
the neutral components of the MSSM Higgs fields)
and
\begin{eqnarray}
I_{1gen}(m_s^2, m_{\tilde{\nu}}^2, m_{\chi_{\alpha}}^2) & =&
\int \frac{d^4 k}{(2 \pi)^4} \frac{1}{k^2 + m_{\tilde{\nu}}^2}
\frac{1}{k^2 + m_{s}^2} \frac{1}{k^2 + m_{\tilde{\nu}}^2}
 \frac{1}{k^2 +  m_{\chi_{\alpha}}^2}  \nonumber \\ & =&
\frac{1}{16 \pi^2} \frac{1}{ \msnu^2 - m_s^2} \left\{
 \frac{1}{m_{\chi_{\alpha}}^2 -\msnu^2 }  -  
\frac{m_{\chi_{\alpha}}^2}{( m_{\chi_{\alpha}}^2 -\msnu^2 )^2} 
    \ln \frac{m_{\chi_{\alpha}}^2}{\msnu^2} \right. \nonumber  \\&&
+ \frac{\msnu^2}{( m_{\chi_{\alpha}}^2 -\msnu^2)( \msnu^2 - m_s^2) } \ln \frac{m_{\chi_{\alpha}}^2}{\msnu^2} \nonumber \\ && \left.
 - \frac{m_s^2}{( m_{\chi_{\alpha}}^2 -m_s^2)( \msnu^2 - m_s^2) } \ln \frac{m_{\chi_{\alpha}}^2}{m_s^2}  \right\}.
\label{I1gen}
\end{eqnarray}

This gives
\begin{eqnarray}
m_{\nu}^{ij}& = & \frac{g^2 B^i B^j~ }{4 \cos^2\beta} 
\sum_{\chi_{\alpha}}   
(Z_{{\alpha}2} - Z_{{\alpha}1} g'/g)^2  m_{\chi_{\alpha}}
 \left\{ I(m_h^2, m_{\tilde{\nu}}^2, m_{\chi_{\alpha}}^2) 
\cos^2(\alpha - \beta) \right. \nonumber \\
 && \left. +  
I(m_H^2, m_{\tilde{\nu}}^2, m_{\chi_{\alpha}}^2) \sin^2(\alpha - \beta) 
- I(m_A^2, m_{\tilde{\nu}}^2, m_{\chi_{\alpha}}^2) \right\},
\label{Imh}
\end{eqnarray}
which does not quite match Grossman and Haber's
result,
but goes to the same limits when various
masses become large. We can check equation (\ref{Imh}) by expanding
the exact result of \cite{DLR}. The result of \cite{DLR}
is for one lepton generation only; we
assume here that the sneutrinos are degenerate,
so we can compare to this calculation.
From \cite{DLR} we have:
\beq
m_{\nu} = \frac{g^2}{64 \pi^2} \sum_{{\alpha}:7..4} m_{\chi_{\alpha}}
(Z_{{\alpha}2} - Z_{{\alpha}1} g'/g)^2
\sum_n(\hat{\nu} \cdot \hat{s}_n)^2  \epsilon_n B_0(0, M^2_n,m^2_{\chi_{\alpha}})
\label{numassexact}
\eeq
where $ \epsilon_n$ is +1 for $s_n $ a CP-even scalar,
and -1 for $n$ a CP-odd. Note that in an $\Rpv$
~theory, the sneutrinos split into a CP-even
and a CP-odd scalar, that are not
mass degenerate. $\hat{\nu}$ is the lepton
direction corresponding to the incident neutrino,
and $ \hat{s}_n$ is the direction in Higgs-slepton
space corresponding to the mass eigenstate $s_n$.
$B_0$ is  a Passarino-Veltman function:
\begin{eqnarray}
 B_0(0, M_s^2,m^2_{\chi}) & 
= & - 16 \pi^2 i \lim_{q \rightarrow 0} \;
\int \frac{ d^{2\omega} k}{(2 \pi)^{2 \omega}}
\frac{1}{[(k+q)^2 - m_{\chi}^2](k^2 - M_s^2)} \nonumber \\
& 
\supset &  - \frac{M_s^2}{M_s^2 - m_{\chi}^2} 
\ln \left( \frac{M_s^2}{ m_{\chi}^2} \right) = I( M_s^2,m^2_{\chi}) ~~~.
\label{B0}
\end{eqnarray}
There are  divergent and scale-dependent
contributions to $B_0$ in addition to 
$I( M_s^2,m^2_{\chi})$; however
these cancel in the sum  over scalars
$s_i$ in equation (\ref{numassexact}).

To compare this result to equation (\ref{Imh}),
we need to expand the  mixing angles 
 $(\hat{\nu} \cdot \hat{s}_n)$
and the masses $ M^2_n$ in $B_i$.
We define the neutrino direction to be 
$ \hat{\nu}_i \equiv \vec{v} \cdot \lambda^i/m_i$
(this is the definition of the charged lepton direction
we have been using all through the paper). 

 As $\Rpv \rightarrow 0 $, 
we make the following identifications:
\beq
\begin{array}{ccc}
s_5 = h_5  \rightarrow h && \\
s_4 = h_4  \rightarrow H &&s_9= A_4  \rightarrow A \nonumber \\
s_3 = h_3  \rightarrow \tilde{\nu}^R_{\tau} && 
s_8= A_3  \rightarrow \tilde{\nu}^I_{\tau} \nonumber\\
s_2 = h_2  \rightarrow  \tilde{\nu}^R_{\mu} &&s_7 = 
A_2  \rightarrow  \tilde{\nu}^I_{\mu}\nonumber \\
s_1 = h_1  \rightarrow  \tilde{\nu}^R_{e} && 
s_6 = A_1  \rightarrow  \tilde{\nu}^I_{e}\nonumber,
\end{array}
\eeq
where $h_n$ is CP-even, $A_n$ is
CP-odd, $ \tilde{\nu}^R$ is the real part of the
sneutrino, and the right hand side of the arrow
 are MSSM fields.

The  diagram that generates a Majorana mass for a neutrino must contain two
units of lepton number violation. There is a contribution
from the mixing angles
 between the neutrinos and  the $s_n, n:9..1 $,
and from the $h_k$ ---$A_k$ mass differences for
$k:3..1$. We consider first the
contributions from  $h_5, h_4$ and $ A_4 $.
The mixing angles
 $(\hat{\nu}_i \cdot \hat{s}_n)$
for $ s_n = h_5, h_4$ or $ A_4 $ will be 
proportional to $B_i$, so $\Rpv$~
corrections to the  $ h_5, h_4$ or $ A_4 $
masses can be neglected in expanding
equation (\ref{numassexact}). They
would be a higher order effect.
We find, for instance,
\beq
\hat{\nu}_i \cdot \hat{h}_5 = \frac{B_i ( \cos \alpha + \
\sin \alpha \tan \beta)}{m_h^2 -   m_{\tilde{\nu}_{i}}^2},
\eeq
which substituted into (\ref{numassexact})
 gives the contribution of the last line of equation (\ref{I1gen})
to the complete result.

The  contribution from $h_3$ and $A_3$ 
( scalars that are mostly  $\tilde{\nu}_{\tau}$)
can be written
\begin{eqnarray}
\frac{g^2}{64 \pi^2} \sum_{{\alpha}:7..4} m_{\chi_{\alpha}}
(Z_{{\alpha}2} - Z_{{\alpha}1} g'/g)^2
\left\{ I( M^2_{h_3},m^2_{\chi_{\alpha}}) - I( M^2_{A_3},m^2_{\chi_{\alpha}})
\right. \nonumber \\
\left.
 - [(\hat{\nu}_{\tau} \cdot \hat{h}_5)^2 + 
(\hat{\nu}_{\tau} \cdot \hat{h}_4)^2 - 
(\hat{\nu}_{\tau} \cdot \hat{A}_4)^2] 
I( M^2_{\tilde{\nu}_{\tau}},m^2_{\chi_{\alpha}}) \right\}.
\label{whatt}
\end{eqnarray}
The first line is the contribution from
the $\tilde{\nu}^R - \tilde{\nu}^I$ mass difference,
and the second line is from the mixing angle:
$(\hat{\nu}_{\tau} \cdot \hat{h}_3)^2 = $$ 
1- (\hat{\nu}_{\tau} \cdot \hat{h}_5)^2$
$ -(\hat{\nu}_{\tau} \cdot \hat{h}_4)^2$.
The third term of equation (\ref{whatt}) gives
contribution from 
the second last term of equation (\ref{I1gen})
to the complete result,
and the first two terms of  
equation (\ref{whatt}) give
the contribution from 
the first two terms  of equation (\ref{I1gen}).
This can be seen  by writing $m^2_{h_3}=
m^2_{\tilde{\nu}_{\tau}} + \Delta m^2_{\tilde{\nu}_{\tau}}/2$,
and 
\begin{eqnarray}
 \Delta m^2_{\tilde{\nu}_{\tau}} & = &
 \left( - \frac{B_3^2(\cos \alpha +
\sin \alpha \tan \beta )^2}{m_h^2 - m_{\tilde{\nu}}^2} - 
  \frac{B_3^2(\sin \alpha -
\cos \alpha \tan \beta )^2}{m_H^2 - m_{\tilde{\nu}}^2} \right. \nonumber \\
&&+ \left. \frac{B_3^2(\cos \beta -
\sin \beta \tan \beta )^2}{m_A^2 - m_{\tilde{\nu}}^2} \right).
\end{eqnarray}
The result in \cite{GH} corresponds to
this contribution.

We can make a similar expansion for  $\hat{\nu}_{\mu} \cdot h_2$ 
and $\hat{\nu}_{\mu} \cdot A_2$, and for $\hat{\nu}_{e}$
assuming as here that there is no flavour violation
amoung the sneutrinos, $e.g.~
(\hat{\nu}_{\tau} \cdot \hat{A}_2) = 0,$ etc.

\section{Discussion}

We now return to our introductory
suggestion that neglecting $\Rpv$~ masses in the loops creates
confusion about basis. Consider the superpotential
for a one-generation model. This could have
two $\Rpv$~ parameters: $\lambda'$ or $\mu_3$. 
As discussed after equation (\ref{simple}), one can define
the Higgs such that $\mu_3 = 0$ or such
that $\lambda' = 0$. In the basis where 
$\mu_3 = 0$, there is the usual loop
neutrino mass corresponding to diagram 1b with $\Rpv$
at the vertices II and VII. Now consider the
basis where $\lambda' = 0$. If the mass
$\mu_3$ is neglected in the loops,
then it appears that there is no loop neutrino
mass. This is  perplexing, because we
know that the tree-level mass is proportional to
the misalignment between the
Higgs-slepton vev and $(\mu_4, \mu_3)$.
If we are in a model where $\vec{\mu} \parallel \vec{v}$,
the basis rotation cannot have moved the
neutrino mass from loop to tree level. Thus,
 it is unclear where did the contribution to the mass  go.
If we include
the loop 1b with $\Rpv$ at points I and VIII,
this confusion is removed. 

We conclude that the basis
in which neutrino mases are calculated is
$irrelevant$, providing that
all the contributions are included.
We believe that ``basis-related'' confusion comes
from neglecting the $\Rpv$~ masses
in computing loop neutrino
masses: the magnitude
of the contribution being neglected
depends on the basis choice.

We therefore sit in  the basis where
the sneutrino has no vev, and consider the
size of different contributions
and the structure of the
neutrino mass matrix in lepton flavour space. 
The tree-level mass contribution to
the neutrino mass matrix is
\beq
[m_{\nu}]_{ij} \sim \frac{\mu_i \mu_j}{m_{\chi}},
\label{2star}
\eeq
 in all cases.

In case A, where we neglect $\Rpv$ masses in the loops, the loop
contributions from the squark/quark and
lepton/slepton loops are respectively of order
\beq
[m_{\nu}]^{ij} \sim 
3 \frac{\lambda^{'iqq}\lambda^{'jqq}(m^d_q)^2}{16 \pi^2 m_{SUSY}}
+\frac{\lambda^{ik \ell}\lambda^{j \ell k}m^e_{\ell}m^e_k}{16 \pi^2 m_{SUSY}}.
\eeq
Note that the quark basis  can be chosen to
simultaneously diagonalise the $\lambda'$ and
down-type Yukawa couplings, but that the lepton
basis that diagonalises the lepton
mass matrix $\lambda^{IJk} v_I$ does
not neccessarily diagonalise 
$\lambda^{i \ell k}$ on indices $\ell k$.

In case B, we include the soft
masses $\{ B_i\}$ and $\{ m^2_{4i} \}$ in
the GH diagram (figure 1c) which contributes
\beq
[m_{\nu}]^{ij} \sim 
\frac{g^2}{64 \pi^2 m_{SUSY}}
( B_i^{\perp} B_j^{\parallel} +  B_i^{\parallel} B_j^{\perp}
 +  B_i^{\perp} B_j^{\perp}) m_{\chi},
\eeq
where $ B_i^{\perp}$ and  $B_j^{\parallel}$
are defined in equation (\ref{14}), and we
have neglected the term of order
$ B_i^{\parallel} B_j^{\parallel}$ because
it is aligned with  the tree-level
mass of equation (\ref{2star}).

In case C, we include all
the $\Rpv$ masses
in perturbation theory. All
the terms listed in column 5 of
table 1 will contribute. Those from diagram
1d, which involves a gauge coupling, are
potentially largest:
\beq
[m_{\nu}]^{ij} \sim 
\frac{g}{16 \pi^2 m_{SUSY}} [(m_e^{i})^2 +(m_e^{j})^2]
\frac{\mu^{i}\mu^{j}}{|\mu|^2}.
\eeq
Note that this is $not$ aligned with the
tree-level mass due to the inclusion of lepton masses.
There are additional contributions of
order $h^4 \mu_i \mu_j/|\mu|^2$ or 
 $h^4 B_i \mu_j/(|\mu| B)$ from mass
insertions on diagrams 1a and 1b.
($h$ is some Yukawa coupling.) These
are listed in table 1.

To summarize 
we have 
estimated one-loop contributions from 
 R-parity violating parameters to the  neutrino mass matrix
in the context of the MSSM. 
We  included the $\Rpv$ bilinears
in the mass insertion approximation, which 
introduces new loops and
new contributions to
the canonical graphs considered in the literature.
For the neutral graph  arising from slepton-Higgs mixing,
our mass insertion results are in agreement with
the perturbative expansion of the exact   result.

\acknowledgments

We would like to thank Gian Giudice for useful conversations.
S.D.  thanks   J. Ferrandis, N. Rius, and W. Porod for
useful conversations and  the
Astroparticle and High Energy Group at Valencia University for 
its warm hospitality while this work was completed.

\end{document}